\documentclass
[twocolumn,showpacs,preprintnumbers,amsmath,amssymb,
superscriptaddress,nofootinbib]
{revtex4-1}

\usepackage{graphicx}
\usepackage{amsmath,bm}

\usepackage{color}

\begin{document}

\title{
%Storm in a Teacup:\\
Dynamics of microdroplets over the surface of hot water
}

\author{Takahiro Umeki}
\affiliation{
Department of Physics, Kyoto University, Kyoto 606-8502, Japan}
\author{Masahiko Ohata}
\affiliation{
Department of Physics, Kyoto University, Kyoto 606-8502, Japan}

\author{Hiizu Nakanishi$^*$}
\email[Correspondence to ]{nakanisi@phys.kyushu-u.ac.jp}
\affiliation{
Department of Physics, Kyushu University, Fukuoka 812-8581, Japan}

\author{Masatoshi Ichikawa}
\affiliation{
Department of Physics, Kyoto University, Kyoto 606-8502, Japan}

\date{\today}

\begin{abstract}
When drinking a cup of coffee under the morning sunshine, you may notice
white membranes of steam floating on the surface of the hot water. They
stay notably close to the surface and appear to almost stick to it.
Although the membranes whiffle because of the air flow of rising steam,
peculiarly fast splitting events occasionally occur.  They resemble
cracking to open slits approximately 1 mm wide in the membranes, and
leave curious patterns.  We studied this phenomenon using a
microscope with a high-speed video camera and found intriguing details:
i) the white membranes consist of fairly monodispersed small droplets of
the order of 10 $\mu\,{\rm m}$; ii) they levitate above the water
surface by 10$\sim$100 $\mu{\rm m}$; iii) the splitting events are a
collective disappearance of the droplets, which propagates as a wave
front of the surface wave with a speed of 1$\sim$2 m/s; and iv) these
events are triggered by a surface disturbance, which results from the
disappearance of a single droplet.
\end{abstract}

\maketitle
%---------------------------------------------------------
%\section*{Introduction}

\begin{figure}
\centerline{\includegraphics[angle=90, width=5cm]
{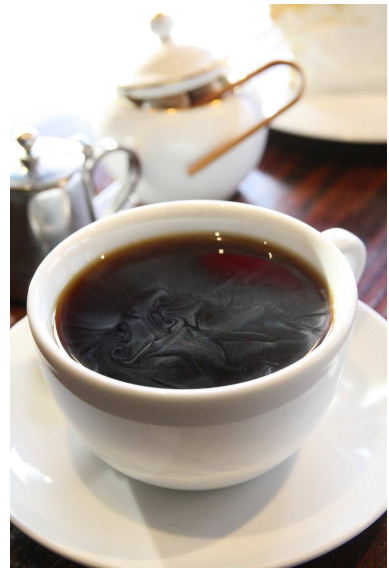} } \caption{
Misty skins over hot coffee.
Peculiar patterns are observed in white membranes on the coffee surface.
The photo is provided through the courtesy of Machida-Sagamihara portal
site Vita (http://vita.tc) and Tamagawa Coffee Club and is allowed for
use in this paper with reference to the companies.
} \label{Photo}
\end{figure}

As a Japanese physicist Torahiko Terada wrote in his fascinating essay
``Chawan no yu (A cup of hot tea)'' in 1922 (in {\it ``The Complete
Works of Terada Torahiko''}, vol 2, pp.3-9, Iwanami, 1997), many
interesting phenomena occur in a teacup filled with hot tea, including
convection, condensation, and vortex flow.  Many are related to a wide
range of natural phenomena in larger scales and have been subjects of
active study in physics during the last few decades.  One phenomenon in
his essay that has not been systematically studied is the misty thin
skin that covers a hot water surface(Fig. \ref{Photo}).  He also
mentioned the crack patterns that run through the skins, and conjectured
that the patterns must be related to the temperature variation developed
by the convection flow in hot water.
Nearly 50 years later, another essay on this phenomenon was written by
Schaefer in 1971~\cite{Schaefer-1971}. He performed some simple
experiments and conjectured that the white skins were made of small
charged droplets that levitated because of the rising evaporation flow
from the surface.  He contended that the patterns on the skins delineate
the Rayleigh-B\'enard convection pattern in hot water and suggested that
the slits suddenly appeared because of micro whirlwinds that develop in
the rising hot moist air flow.
%-----------------
%-----------------

Recently, a similar phenomenon was accidentally discovered by Fedorets
in a different setting when he was examining the surface of water layer
on an ebonite substrate heated with a lamp~\cite{Fedorets-2004}.  He
found a layer of microdroplets of approximately 10 $\mu{\rm m}$ forming
a triangular lattice and observed that a segment of a droplet cluster
suddenly disappears within 0.04 sec, i.e., a single-frame interval of
his video camera.  In a series of
works~\cite{Fedorets-2004,Fedorets-2005,Arinshtein-2010,Fedorets-2011},
Fedorets and co-workers examined several conceivable mechanisms of the
droplet levitation and suggested two possibilities: the force from
droplet spinning because of the Marangoni effect and the Stokes drag
force because of the rising evaporation flow.

In this work, we examine the surface of bulk hot water using a
microscope with a high-speed video camera, and report what the white
membranes are and how the ``cracking events'' actually occur.

\section*{Results}
%\subsection*{Experiments}

We performed two sets of experiments: preliminary observations on
several types of hot water/beverages and detailed observations on
hot tap water with a high-speed video camera.
The preliminary experiments were performed with a simple setting; the
hot water/beverage in a beaker was examined by eye and with a video
camera.
%a microscope (KEYENCE VH-Z75, Japan) 
%with a video camera of the frame rate 30 fps.  
%
We made observations on several types of hot water, such as coffee, tea,
water with detergent, and pure water filtered using a Milli-Q system,
and we did not find significant differences in the white membranes for
all cases.
The microscope observations showed that the white membranes consist of
micro droplets of the order of 10 $\mu\rm m$.
The droplets were also found on the surface of moderately hot water with
a temperature of approximately 50$^\circ$C, although they were notably
sparse.

\begin{figure}
%\centerline{\includegraphics[width=8cm]{./Fig-1-rev-crs-hn.eps3}
\centerline{\includegraphics[width=8cm]{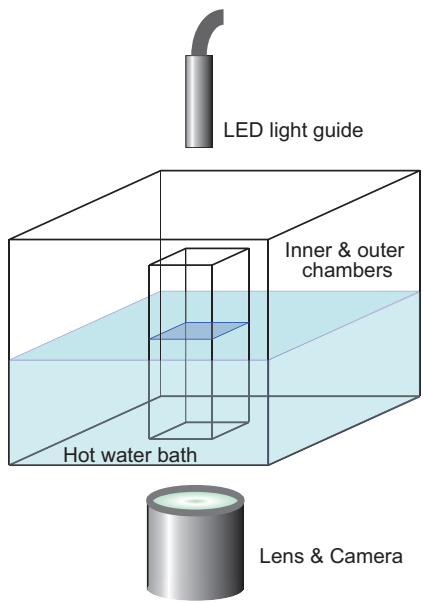}
}
\caption{
Schematic diagram of the experimental setup.}
\label{Setup}
\end{figure}

After the preliminary observations, we made closer observations of hot
tap water with a high-speed video camera using a setup designed
for this phenomenon (Fig. \ref{Setup}). The experimental setup had
a water container, a liquid light-guide illumination system,
and a microscope with a high speed-video camera (KEYENCE VW-9000,
Japan).
To avoid water condensation on the lens,
the water surface was observed from below
under bright- or dark-field illuminations from above. 
The dual wall chamber was used for the water container to reduce
external disturbances.  The container was filled with hot water of 60
$\sim$ 90$^\circ$C in the room temperature environment of 20 $\sim$
25$^\circ$C; the room temperature and humidity were not controlled
during the experiments.  The recorded images were analysed using 
image-processing software, mainly ImageJ.

%---------------------------------------------------
%\section*{Results}

%\paragraph*{Observations by a simple setting}

\begin{figure*}
% \center{\includegraphics[width=8cm]{./SnapShot-2.eps3}}
 \center{
\includegraphics[width=15cm]{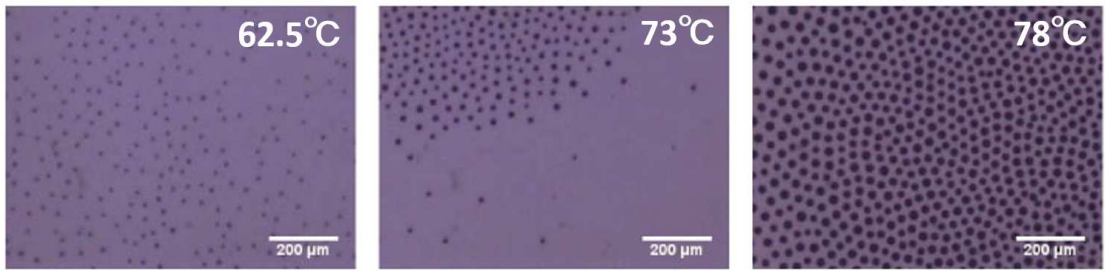}
}
\caption{
Microscope images of the hot-water surface
at the temperatures $T=62.5$, 73, and 78$^\circ$C. 
} \label{SnapShot-2}
\end{figure*}

\vskip 2ex\noindent{\bf Snapshots.}
Figure \ref{SnapShot-2} shows typical microscope pictures at
three different temperatures.  Droplets are scattered on the surface and
drifting together at approximately a few mm/s.
They do not stay on the surface but levitate above the surface, which
is confirmed with the independence of their drift motions on those of
dust particles that float on the surface. The distance above the surface
is estimated to be 10 $\sim$ 100 $\mu\rm m$ from the focus depth of the
microscope.

Figure \ref{SnapShot-2} shows the following:
that the droplets are more populous and larger at higher temperature;
that the droplets are relatively monodispersed at each temperature; that
there is a clear boundary between the dense and the sparse regions of
the droplets; and that the droplets tend to form a triangular lattice
in the dense regions.

\begin{figure}[b]
% \center{ \includegraphics[width=7cm]{./size-dist.eps} } 
 \center{ \includegraphics[width=7cm]{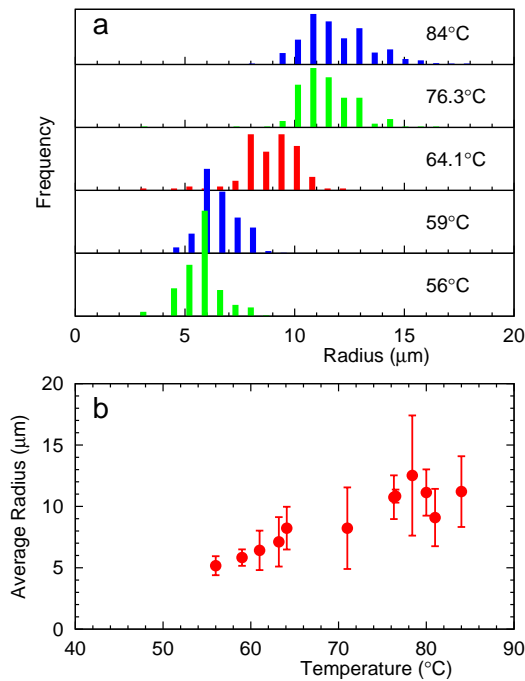} } 
\caption{
Size distribution (a) and average size of the droplets (b)
at various temperatures.  The error bars for the average radii
represent the standard deviation of the distributions.  The data set for
each temperature contains several hundreds of data points of droplet
radii.
}
\label{size-dist}
\end{figure}

Figure \ref{size-dist} shows the size distributions and average sizes of
the droplets at several temperatures. The radius distributions show
narrow dispersion, whose standard deviations are typically less than
half of their average value.  The average radius of the droplet
gradually decreases when the temperature decreases. The data at
$T<60^\circ$C may contain systematic error because the average radius
becomes comparable to the resolution limit of the images, i.e., 4.74
$\mu$m for 1 pixel.

%--------------------------
%--------------------------

%---------------------------------------------
\vskip 2ex\noindent{\bf Observations with high-speed camera.}

Careful examination of the video images reveals that the droplets
occasionally fall from above to settle above the surface, and they
occasionally fall down to the water surface to individually disappear.
In addition to such {\it individual disappearances}, we observed
occasional {\it collective disappearances} of hundreds of droplets
within a single-frame interval of the 30 fps video camera; this
collective event corresponds to the aforementioned splitting event.
To study this fast process during the collective disappearance, we made
observations using the high-speed video camera with frame speeds of
1000 and 8000 fps and a shutter speed of 1/16000 sec.

\begin{figure*}
% \center{\includegraphics[width=8cm]{./CollectiveVanish-2.eps3}}
% \center{\includegraphics[width=8cm]{./Figure-3-Nakanishi.eps}}
 \center{\includegraphics[width=12cm]{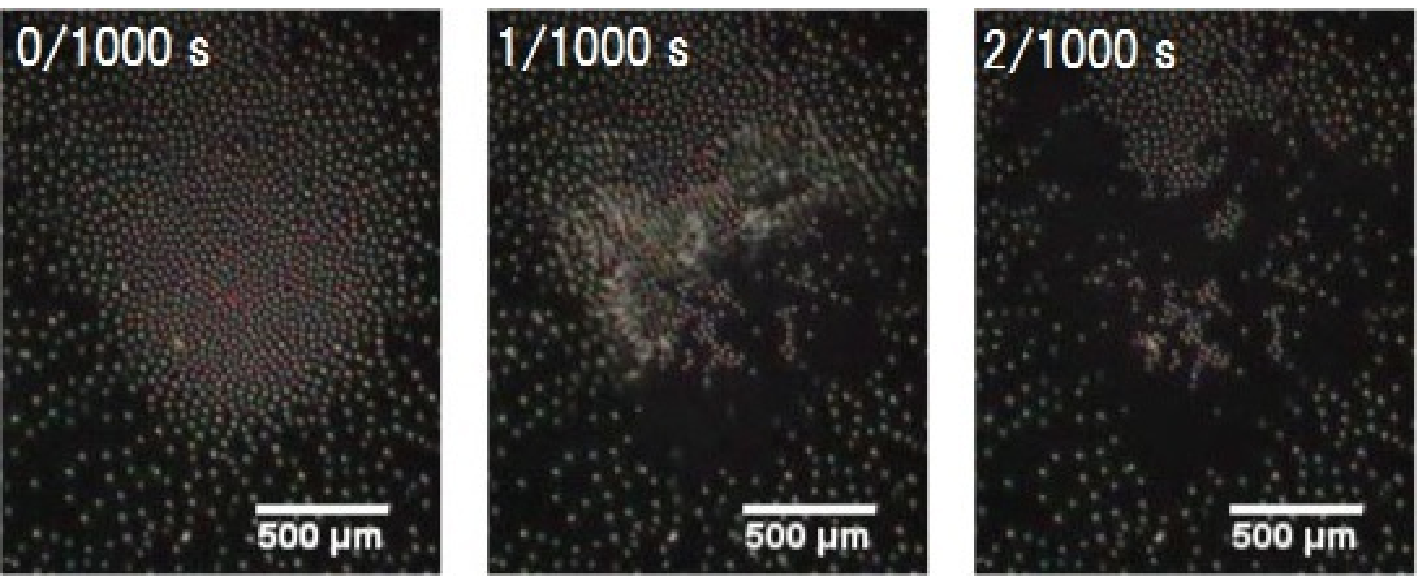}}
\caption{
Collective vanishing event of a droplet cluster.  Three
consecutive frames of the 1000 fps video are shown.  Levitating water
droplets are observed as the bright spots under dark-field illumination.
(Supplementary video 2)
}
\label{CollectiveVanish-1}
\end{figure*}

Figure \ref{CollectiveVanish-1} shows three consecutive frames of 1000
fps video, which captured a collective vanishing event. A
millimetre-sized droplet cluster vanished within a couple of frame
intervals. The streaks appeared in the middle frame because of the
droplet motion during a single-frame exposure.  It should be noted that
not all droplets in the region vanished, and there were {\it surviving
droplets}, i.e., some isolated droplets and/or clusters of droplets
survived in the region that had a collective event.
The videos also show a continuous falling of droplets from above onto
the surface, some of which settled in the clusters and caused
rearrangement of the surrounding droplets (Supplementary video 1).
%--------------------------
%--------------------------

\begin{figure}[b]
% \center{\includegraphics[width=8cm]{./Fig-4-rev.eps}}
 \center{\includegraphics[width=8cm]{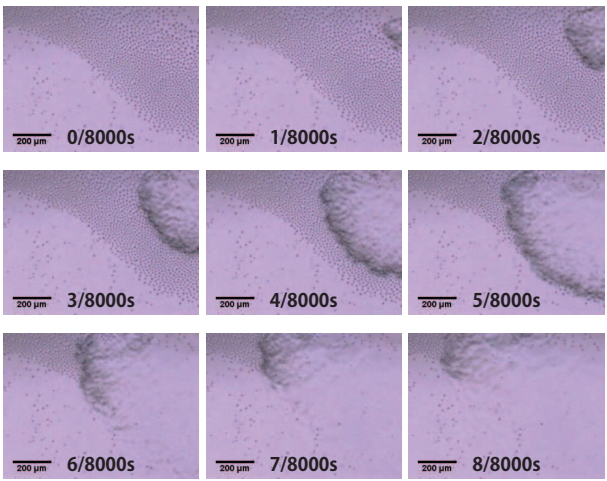}}
\caption{Wave front propagation of a collective vanishing event.
Nine consecutive video frames of 8000 fps are shown.
(Supplementary video 3)
}
\label{Wave-1}
\end{figure}

\begin{figure}[b]
% \center{\includegraphics[width=7cm]{./Trigger-arrow.eps3}}
 \center{\includegraphics[width=7cm]{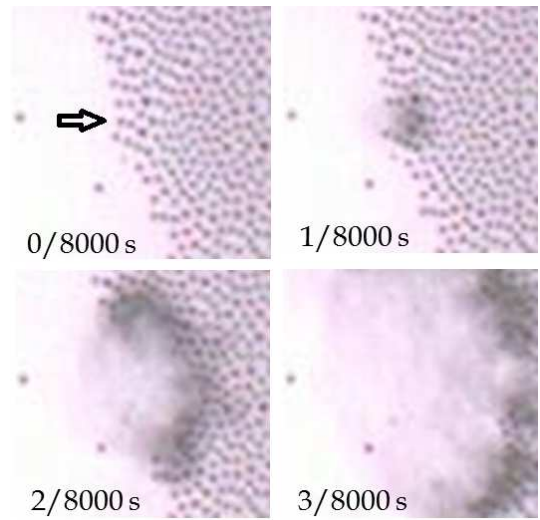}}
\caption{
Initial stage of a collective vanishing event.  The
enlarged images from the high-speed video frames show that the
collective vanishing event is triggered by a disturbance, which is
caused by the disappearance of a single droplet, as indicated by the
arrow.
}  \label{Trigger}
\end{figure}

Figure \ref{Wave-1} shows nine consecutive frames of the 8000 fps video,
which reveal that the vanishing event propagated as a wave front at
speed of approximately 1m/s. It is also observed that the wave front was
accompanied by a surface wave of the water, which propagated at an
approximately identical speed.
The four consecutive frames in Fig. \ref{Trigger} capture the initial
stage of a collective event.  The collective event
was triggered by the disappearance of a single droplet, from which the
disturbance wave front propagated outward and  caused a massive
disappearance event that involved thousands of droplets.
We also observed {\it concurrent events}, where two collective and/or
individual events occurred within a 10 ms time interval, although they
did not seem causally related through the surface wave because they were
spatially separated from each other.

\begin{figure*}
%\centerline{\includegraphics[width=10cm]{./Fig-detergent-78C.eps3}}
\centerline{\includegraphics[width=15cm]{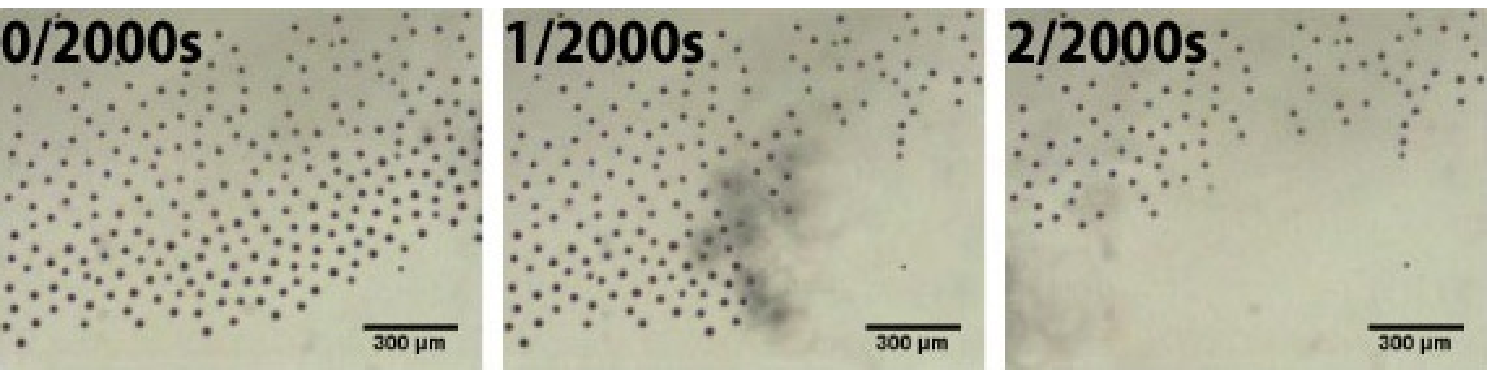}}

\caption{
Vanishing event on the hot water with detergent.  Triton X-100
was dispersed with a concentration of 0.3 mM, which is above the
critical micelle concentration at room temperature, i.e., 0.24 mM.  The
water temperature was 78$^\circ$C.
}  \label{detergent}
\end{figure*}

Finally, Fig. \ref{detergent} shows a vanishing event in the system with
the surfactant Triton X-100.  The concentration was 0.3 mM, which is
above the critical micelle concentration at room temperature, i.e., 0.24
mM.  One can observe that the event is qualitatively similar to the one
in the tap water.  The surfactant does not appear to significantly
change the droplet size and the propagation speed.

%--------------------------
%--------------------------
%--------------------------

%---------------------------------------------
\section*{Discussion}

With these observations, we found the following:
1) The phenomenon is not sensitive to the type of water, i.e., coffee,
tea, water with detergent, and Milli-Q water;
2) the white membranes consist of the fairly monodispersed microdroplets
that fall from above and levitate immediately above the water surface;
3) the typical size of the droplets is on the order of 10 $\mu{\rm
m}$, the average size decreases as the water temperature decreases, and
the levitation height is 10$\sim$100 $\mu{\rm m}$;
4) the splitting events in the membranes are the propagation of the
collective disappearance of droplets, which is triggered by a single
droplet; and
5) their propagation fronts are accompanied by the surface wave, 
both of which propagate at ca. 1m/s.

These observations provide more questions than answers:
1) What is the levitation force?
2) Why are the droplets so monodispersed?
3) What triggers the collective vanishing event?
4) How does the vanishing wave front propagate?
5) Why do some droplets survive after a vanishing event?

%----------------------------------------------------------
\vskip 2ex\noindent{\bf Non-coalescent drops.}
Before discussing some of these issues, let us briefly mention an
apparently similar phenomenon, i.e.,  non-coalescent drops, which
you may observe when you make drip coffee.
Some drops stay on the coffee surface for several seconds and roll over
it before they coalesce into the coffee in the pot. The droplets float
above the surface also in this case, but the droplet size is notably
different: 1 cm for the dripping droplets and 10 $\mu$m for the droplets
in the white membranes.  This type of non-coalescent liquid drops has
been studied in various situations\cite{Rayleigh-1899,Neitzel-2002}, and
it has been shown that two surfaces of the same liquid that are pressed
against each other can be stabilised without coalescing for quite a long
time because of a thin air layer, which is maintained by various
mechanisms, such as the air drag by liquid viscosity\cite{Couder-2005},
externally driven air flow\cite{Sreenivas-1999}, air flow driven by the
thermal Marangoni
effect\cite{DellAversana-1996,DellAversana-1997,Monti-1998,Savino-2003},
surfactant surface elasticity\cite{Amarouchene-2001}, or
vibration\cite{Couder-2005}, etc.  However, the relevance to the present
phenomenon is not direct because of the size difference.

%----------------------------------------------------------
\vskip 2ex\noindent{\bf Levitation force.}
The required force to levitate a 10$\mu{\rm m}$ diameter droplet against
gravitation is approximately 5pN.  The levitation force may depend on
the water temperature because the average droplet size increases with
the water temperature.
 The evaporation flow from the hot water surface is a candidate as
Schaefer previously conjectured in his essay \cite{Schaefer-1971}.
Recently, this mechanism was examined in more detail by Fedorets and
co-workers\cite{Fedorets-2005,Fedorets-2011}.  They also explored
another mechanism; the droplets may be levitated by the air flow induced
by the spinning motion because of the Marangoni
effect\cite{Fedorets-2005}.
%
%the Marangoni effect may cause spinning motion of the
%droplets and the air flow induced by the spinning motion make the
%droplets levitate\cite{Fedorets-2005}.
%
They estimated that the droplets could spin as fast as 50 rotations per
second (rps) by a possible temperature variation in the droplets.  Both
mechanisms should be sensitive to the water temperature and
environmental humidity.
Another possibility is an electrostatic force because small droplets can
be easily charged. In the present case, the droplets are likely charged
because they form a triangular lattice in densely populated regions.
With this mechanism, the levitation force should be sensitive to the
electric status of the surface, which should be affected by  salt or
other impurities in the water.
It is puzzling that the phenomenon appears insensitive to variations in
the water temperature, humidity, and water purity.

\vskip 2ex\noindent{\bf Size distribution of the droplets.}
%Monodispersity.}
%
The droplets are relatively monodispersed.  The standard
deviations of the linear size distributions are less than 50\% of the
averages.
If there is no particular size of approximately 10$\mu{\rm m}$ for the
droplets to be thermodynamically stable, their size may be selected
based on the force balance between gravitation and the levitation
force.

In a slightly different situation from the present case, Arinshtein and
Fedorets made an interesting observation on this
connection\cite{Arinshtein-2010}: there appeared droplets as large as
100$\mu{\rm m}$ in diameter over the water layer on the ebonite
substrate when the water surface was heated at a localized region of 1
mm in size.
%

%--------------------------
\vskip 2ex\noindent{\bf Vanishing events.}
Droplets individually and collectively vanish.  In a collective event,
the vanishing front propagates at approximately 1m/s, which is close to
the capillary surface wave speed of the water with a wave-length of
0.1$\sim$1 mm. Thus, it is natural to assume that the droplets are
swamped by the surface wave, which is sustained by the surface energy of
the swamped droplets.  A simple estimate of the energy balance shows
that the energy input from the swamped droplets can maintain the surface
wave with a sufficiently large amplitude to swamp the droplets in the
regions of high density of droplets (see Appendix section).

%-----------
We observed that a single-droplet disappearance caused a collective
event, although there are also many isolated events that do not
trigger a collective event.
%
%It seems that 
The single-droplet events tend to occur in less populous regions, which
is consistent with the above mechanism of collective vanishing because
in a sparse region, the disturbance decays before it reaches
neighbouring droplets; otherwise, it should trigger a collective
vanishing event.

Whether a vanishing event triggers a collective event or not, it is not
clear what causes a droplet to vanish in the first place; the reason may
be fluctuations in the evaporation flow or an electric disturbance that
is caused by a cosmic ray if the droplets are levitated by the
electrostatic force.
The observed concurrent events are likely to be triggered by a common
disturbance in the levitation force because it is highly improbable that
more than one event occurs in less than 10 ms in a small area of the
microscope field of view.
%------

If the levitation force is provided by the temperature, the spatial
pattern of the droplet disappearance events may reflect spatial
variations in the surface temperature of the water, which would explain
the observation of Terada and Schaefer\cite{Schaefer-1971} that the
splits in the white membranes appear to delineate the Rayleigh-B\'enard
convection pattern in the cup.  The collective disappearance events
should in the low-surface-temperature region of descending
convection, where the levitation force is weaker and thus the
droplets stay closer to the surface.

%------
%--------------------------
\vskip 2ex\noindent{\bf Surviving droplets.}
If the swamp mechanism of the collective droplet disappearance is true,
some droplets can survive if they are levitated higher.  However, the
surviving droplets do not necessarily appear smaller than the vanished
droplets, which does not appear consistent with the idea that the
levitation height is determined by the force balance because a smaller
droplet should stay higher in this mechanism.
To determine the mechanism, experimental observations of
the levitation heights and their correlation with the droplet size and
surface temperature are important. 

%------
%-------------------------------
%\paragraph{Concluding remark:}

\vskip 2ex

In conclusion, a cup of hot tea continues to provide us with interesting
phenomena that deserve scientific research.  The phenomenon that we
studied here can be observed everyday and should have been noticed
by many scientists, yet very few people appear to have imagined such
fascinating phenomena happening in a teacup.

%-----------------------------------------------------------------------
\appendix
\section*{Appendix}

In this section, the amplitude of the surface wave is estimated based on
the described swamping mechanism.  In the steady propagation, there
should be an energy balance between the energy input from the swamped
droplets and the energy dissipation in the surface wave because of the
viscosity.  Assuming that the surface wave is localised around the wave
front region with the width and depth of the wave-length $\lambda$, the
energy dissipation per unit time per unit length along the wave front is
estimated as
\begin{equation}
e_{\rm diss} \sim 
\eta \left({h\over\tau\lambda}\right)^2 \times \lambda^2 ,
\end{equation}
where $\eta$ is the water viscosity, $\tau$ is the wave period,
and $h$ is the height or amplitude of the wave.
The energy input is estimated as the product of
the number of swamped droplets and the surface energy of a droplet,
\begin{equation}
 e_{\rm input} \sim vn \times \sigma d^2 ,
\end{equation}
where $v\equiv \lambda/\tau$, $n$, $\sigma$, and $d$ are the wave speed,
area density of the droplets, surface tension of the water, and droplet
diameter, respectively.  By equating $e_{\rm diss}$ and $e_{\rm input}$,
we obtain
\begin{equation}
 h\sim \left(\sqrt{{1\over\eta}\tau^2 v n \sigma}\right) \times d.
\label{height}
\end{equation}
If we use
\begin{align}
 \lambda\sim 100\mu{\rm m},\quad
 v\sim 1{\rm m/s},\quad
 n\sim 1/(30 \mu{\rm m})^2
\end{align}
with the following viscosity and the
surface tension for water,
\begin{equation}
  \eta\sim 10^{-3}{\rm Pa\cdot s}, \quad
  \sigma\sim 7\times 10^{-2}{\rm N/m},
\end{equation}
Eq.(\ref{height}) is estimated as
\begin{equation}
 h \sim 30\times d.
\end{equation}
This estimate suggests that the sufficiently large amplitude of the
surface wave can be maintained by the surface energy of the swamped
droplets for the observed droplet density in densely populated regions.

%--------------------------------
\begin{figure}
%\centerline{\includegraphics[width=6cm]{./droplets-surface_wave-3.eps3}}
%\centerline{\includegraphics[width=6cm]{./Figure-9-Nakanishi.eps}}
\centerline{\includegraphics[width=6cm]{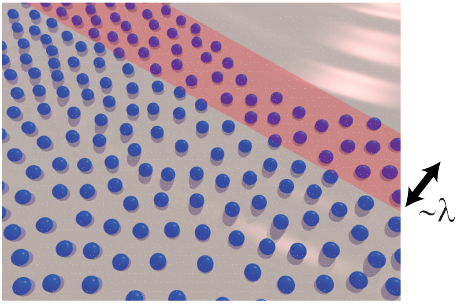}}
\caption{
Cluster of droplets being swamped by surface wave front. The
shaded area is the wave front region of the wave length width.
}
\end{figure}
%--------------------------------

%==================================================================
%%\bibliography{../references}

%=====================================================-
%------------------------------------------------
%-------------------------------
\section*{Acknowledgement:}
The authors thank Clive Ellegaards, Namiko Mitarai, Takahiro Sakaue, and
Yoko Ishii for stimulating discussions.  This work was supported by JSPS
KAKENHI Grant Number 25610124.

%-------------------------------
\end{document}